# Local Core Members Aided Community Structure Detection

Xiaoping Fan [1,2] Zhijie Chen [2] Fei Cai [3] Jinsong Wu [4] Shengzong Liu[2] Zhining Liao[5] Zhifang Liao [3,*]

(1) School of Information Science and Engineering, Central South University, China
(2) Information Management Department, Hunan University of Finance and Economics, Changsha, Hunan, China
1. (3) School of Information Science and Engineering, Central South University, China
(4) Department of Electrical Engineering, Universidad de Chile, Santiago, Chile
(5) Division of Health & Social Care Research, Faculty of Life Sciences & Medicine, King's College London, UK
Correspondence Author: zfliao@csu.edu.cn

**Abstract**: The relationship of friends in social networks can be strong or weak. Some research works have shown that a close relationship between friends conducts good community structure. Based on this result, we propose an effective method in detecting community structure in social networks based on the closeness of relations among neighbors. This method calculates the gravity between each neighbor node to core nodes, then makes judgement if the node should be classified in the community or not, and finally form the process of community detection. The experimental results show that the proposed method can mine the social structure efficiently with a low computational complexity.

# 1 Introduction

Community discovery is one of the core research areas in social network analysis [1] which is also benefit to other related research areas, such as public opinion monitoring, advertising precision delivery, impact analysis. As early as in 1927, Stuart Rice [2] proposed the discovery of small political groups based on voting patterns. Until 2002, Newman et al. proposed the GN algorithm [2], and community analysis began to flourish. The GN algorithm is a classical graph splitting algorithm, which uses the edge betweenness to measure the importance of the edge in the entire network, the larger betweenness the edge has, the edge is more likely to connect two different communities. Newman proposed a fast agglomeration algorithm based on the GN algorithm[3] via repeatedly calculating the network shortest path to update the edge betweennesses until the network is divided into appropriate community structures, for an unweighted graph with n nodes and m edges, the time complexity is as high as $(n^3)$. Through continuously merging nodes and communities, the highest modularity of the community is generated. The efficiency of the algorithm is greatly improved. In order to further reduce the computational complexity, many improved algorithms, such as CNM algorithm [4] and FastUnfolding algorithm [5] based on modularity optimization, have largely sacrificed the quality of the results in order to ensure the speed of computation. Rosvall et al. [6] proposed the Infomap algorithm, a method based on information theory and random walks, which uses Huffman coding optimized by the code length to encode the traveling path and the objective function. Raghavan et al. [7] proposed a Label Propagation Algorithm (LPA) based on the idea of message passing. Each member keeps transmitting the label information he or she obtains, and according to the most frequent occurrences in his or her neighbor tags to update their own labels, and finally the nodes with the same label is divided into the same community. The computational complexity of the LPA algorithm is $(km)$, so the LPA algorithm can deal with a large-scale network efficiently. But the algorithm does not have a unique community division, and the quality of community partitioning is lower in low-density networks. Beyond LPA algorithm, there have been many improved algorithms, such as HANP algorithm[8], SLPA algorithm[9] and BMLPA algorithm[10]. In addition, there have been many research works using probability models to generate the target networks for community discovery issues. Yang et al. [11] proposed a community discovery algorithm (BigCLAM method) based on the node-community affiliation. By designing the probability function of the connection between nodes and using the EM algorithm to iterate the membership value of the node to the community until the algorithm converges, the algorithm is suitable for large-scale networks with millions of nodes, but it is necessary to change the value of *K* (the number of communities) and repeat the calculation, and get the best number of communities. Wang et al. [12] utilized multiple social features of the nodes' behaviors to quantify the nodes pairs' social links and these social links can be used to construct the friendship communities of the nodes. The algorithm shortens the routing delay and increases the successful delivery ratio, thereby improving the routing efficiency. Wang et al. [13] presented a trustworthy crowdsourcing model in SIoT(Social Internet of Things), social cloud provides compute and storage functions, and works as a service provider to bridge end users and sensing entities; sensing entities receive tasks and rewards from a service provider and feedback data. Then they incorporated a reputation-based auction mechanism into crowdsourcing to perform winner selection and payment determination by evaluating the reliability of crowdsourcing participants. However the algorithm mainly focus on availability threat issues relevant to designing trustworthy crowdsourcing in SIoT.

In the process of community discovery, most of the previous algorithms have used modularity as the optimization function to divide the community, or measure the node's importance by the degree of nodes to discover the community[14]. These methods neglect the local community characteristics of neighborhood sets in the network, and the close neighborhood sets also can produce good community structure [14]. Meanwhile the core nodes in the community structure can affect the community structure and guide the community development trend. Under this consideration, in this paper, we propose a local core node-based diffusion method to discover the community structure in the network via analyzing the neighborhoods community importance of nodes in local networks and using the conductivity to measure the importance of nodes. The structure of the paper is as follows: Section 1 introduces the relevant content of community structure, and briefly discusses several community discovery methods; Section 2 gives the relevant definition and algorithm model of community discovery; Section 3 is the experimental results and analysis; Section 4 summarizes the full text.

## 2 NCB Algorithm

In the real world, community always appears around some influential nodes of the social network at the beginning, so we can use these nodes and their neighborhoods to simulate the progress of community's appearance in order to explore the structure of social network. According to power-law distribution of the node degree, we experientially conclude three rules as follow:

(1) Some nodes in a network are more important than others, and these nodes build connections between nodes with their significant influence, which contributes to the appearance of community. Thus, the key nodes in community play a decisive role in maintaining the stability of the community.
(2) Community cannot be star-like structure, so the node with big degree may not be the core node. And nodes in the network follow the Long Tail Effect and the ternary closure. We assume that there are four people in the network, C is a friend of A and B, but A and B are strangers, D is a stranger of A, B and C. Then, A is more likely to be a friend of B than D.
(3) The key nodes will appear in community, and they are not supposed to appear on the bridge which connects different communities.

According to above rules, we proposed NCB(Community Detection Algorithm based on Core Members in Neighbourhood), a new model to describe the progress of community's appearance. NCB has three stages, including original, extension and update, which are shown in Figure 1.

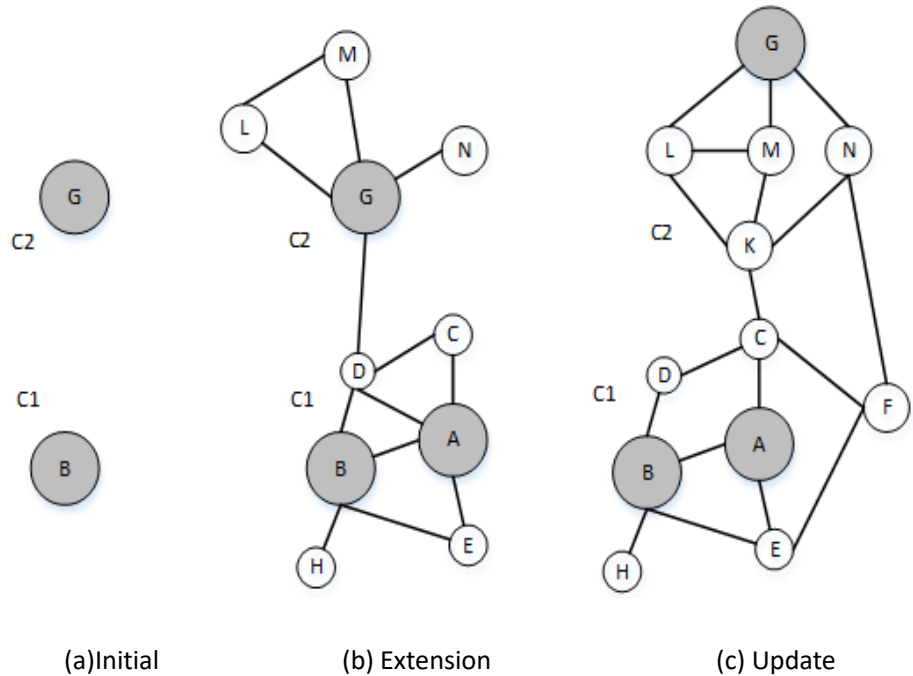

(a) Initial      (b) Extension      (c) Update

Figure 1 Three Stages of NCB

Initial stage: Find all the key nodes, and merge closer nodes into a community according to the connections between key nodes. For instance, nodes B and G are chosen to be the key nodes of community because they are most infective.

Extension stage: Merge the direct neighbors of key nodes into community, and judge the degree of other nodes that are not directly connected to key node whether these nodes join in this community. In Figure 1, nodes C, E and H joined in community $C_1$, and nodes L, M and N joined in community $C_2$. D joined in $C_1$ because D has a tighter connection with $C_1$.

Update stage: Choose the nodes from the expanded neighbor set of key nodes, merge the nodes that are most likely to join the current community into the network. As shown in Figure 1, node K joins in community $C_1$ and node F joins in community $C_2$. Expand the cover area from the center during all the process, if the node who has not merged into any community, we compute the closeness between the nodes of each community.

NCB stimulates community's appearance via analyzing how closely a node connects to community. Three key questions in the stimulation progress are raised as follow:
(1) How to find the key nodes in the network ?
(2) How to judge which community the other common nodes want to join in ?
(3) How to judge whether a community accepts a node ?

**2.1 The node importance calculation**

Given a node set $S$, the set $\bar{S}$ is a complement of $S$, $V = \bar{S} + S$. For any disjoint of two node sets, such as $S$ and $T$, $E(S,T)$ represents the edges between set $S$ and $T$, $cut(S)$ represents the size of the partition produced by the $|E(S,T)|$, which is the number of edges between two node sets.

We randomly select a node to start a random walk test. The conductivity is the probability that the point goes in and out of the node set, let $d(S)$ to represents the summation of degrees of all nodes in set $S$, and let $edges(S)$ to represent twice of the total number of edges, The following expression can be obtained.

$$edges(S) = d(S) - cut(S) \qquad (1)$$

$\phi(S)$ is the conductivity of the node set S, then $\phi(S)$ can be expressed as follows,

$$\phi(S) = \frac{cut(S)}{\min(d(S), d(\bar{S}))} \in [0,1] \qquad (2)$$

The conductivity is measured by the smallest values of the set $S$ and $\bar{S}$, $\phi(S)$ is the probability that an edge can be selected from a smaller set to cross the segmentation. Let the neighbor set of node $v$ is $(v)$, then smaller $\phi(N(v))$ means closer the node $v$ with its neighbors, the more obvious the local community characteristics.

**2.2 Notions**

**Definition 1. Vertex Neighborhood**. The neighborhood of a node can be defined as: $N_1(v) = \{w | d(w,v) = 1\}$, and the neighborhood of a community can be described as: $N_1(C) = \{w | d(w,C) = 1\}$. $d(w,v)$ stands for the shortest distance from node $w$ to $v$, and $d(w,C)$ represents the shortest distance between $w$ and the community $C$.

Take each node and the neighborhood of the node as a cut, and define the cut as a neighborhood community. If the neighborhood of a node has a less conductance score than all other neighbors, then the neighborhood is said to be a best local community. That is, each of the best local community to meet the following condition,

$$\forall w \in N_1(v), \phi(N_1(v)) \leq \phi(N_1(w)) \qquad (3)$$

where $N_1(v)$ stands for a set of neighbor nodes $v$ with a distance of 1.

According to the definition of conductance and the best local community, smaller the conductivity of the local minimum community is, more obvious the community structure is, and the core node in the best local community can be regarded as the initial community.

**Definition 2. Community Gravitation**. When studying the relationship between the node and the community, we can refer to the law of gravitation, take the community as a star and the node as an asteroid, the number of common neighbors between the community and the node can be used as an element to calculate the magnitude of the gravity between them. The greater the gravity is for $v$, the more desire to join the community for $v$. As $v$ perhaps will be accepted the gravitation from multiple communities, it can be normalized by the degree ($d_u$) of $v$ for gravitation attracted of $v$ as follows,.

$$CF_u(C) = \frac{e_C^u}{d_u} \tag{4}$$

where $e_C^u$ is the number of edges from $u$ to $C$, and $d_u$ is the degree of node $u$.

**Definition 3. Community Stability**. The stability of a community is determined by the connection degree of its internal membership, the closer the membership is, the more stable the community is. Thus, the ratio of the twice number of edges within the community to the sum of degrees of the nodes can be used as the stability of the community.

$$s(C) = \frac{2 \cdot e_C^{in}}{d(C)} \tag{5}$$

where $e_C^{in}$ represents the number of edges in the community $C$, $d(C)$ stands for the sum of degrees of all nodes in the community . In the social network, the size of the community is a small part relative to the entire network, that means that $(S) \ll d(\bar{S})$, thus $s(C) = 1 - conductance(C)$ .

**Definition 4 Capture Factor**. If the community $C$ becomes more stable after trapping the node $v$, then $C$ accepts that the node $v$ joins in the community, otherwise $C$ does not accepts

$$\varepsilon = s(C \cup \{v\}) - s(C) \tag{6}$$

If $> 0$ , node $v$ will increase the stability of the community, otherwise it will weaken.

**2.3 The Description of Algorithm**

NCB algorithm is a kind of heuristic community detection algorithm based on local optimal neighbor set. Unlike other heuristic community detection algorithms, NCB algorithm does not select the degree of nodes as the seeds of community, but the nodes with the best conductivity as the core node. Then, the algorithm applies the community's gravitation to the node to measure whether the node is willing to join the community and adopts the community's stability as the capture factor to judge whether the community accepts the node.

The pseudo-code of the algorithm is as follow:

**Input**：The original Network(nodes and edges)

```
Output: Community structure
1: put all the nodes into the List
2: For  each node in List do
3:    Get the node's neighborhoods
4:    Compute conductive
5:    Find the smallest conductive node and neighborhood
6: End for
7: For  each node  not in   do
8:    select the node with the largest gravitation as candidate node
9:    If the stability of community
10:       accepts the node
11: End For
12: Return
```

**2.4 The Analysis of Algorithm**

NCB algorithm applies a heuristic approach, which starts the search in the neighbor nodes set of the core nodes. The computational complexity for each output of a community depends only on the size of the community's neighbor set, not the size of the entire network. The analysis on the computational complexity of each step is provided as follows.

We define the total number of nodes in the network is $N$, the total number of edges is $E$, the average degree of nodes is $m$, the number of communities is $c$, and the average number of nodes in the community is $k$.

(1) The computational complexity on the process of node conductivity calculation is $O(Nm)$;

(2) The computational complexity on constructing seed sequences by Minimum Heap is $O(N \log N)$;

(3) The average time spent in constructing the initial community and solving the gravitation of a neighbor $(c \cdot m)$ ;

(4) The total time complexity on the process of iterating the community's expansion and using the binary sort tree to update the node's neighbors and their gravitational value is $O(k \cdot m \cdot \log(k \cdot m))$. Owing to $c \cdot k \cdot m = E$, $k \cdot m = E/c$, $O(c \cdot k \cdot m \cdot \log(k \cdot m)) = O(E \log(E) - E \log(c))$, the time complexity of the NCB algorithm is $O(E \log(E))$. Moreover, the NCB algorithm uses the core node to iteratively calculate in the local area, and the community size increases gradually, so $O(E \log(E))$ is only the complexity of the worst case.

# 3 Experiments

## 3.1 The dataset

In this experiment, we explore two different genres of data: (1) Dataset with real community partition, it includes Karate Club Network and Dolphin SocialNetwork; (2) Real network dataset without community from Cond-matscientific collaboration network, Twitter network, Brightkite network. Detailed descriptions are shown in Table 1.

Table 1 The Data set

| Dataset | Node message | Edge message | Node number | Edge number |
|---|---|---|---|---|
| Karate | member | communication | 34 | 78 |
| Dolphins | dolphin | communication | 62 | 159 |
| Cond-mat | author | cooperation | 40,421 | 175,692 |
| Twitter | user | follow | 23,370 | 33,101 |
| Brightkite | user | friends | 58,228 | 214,078 |

## 3.2 Result Analysis

There exist obvious long tail effects in the social network, and degree of the node obeys the power law distribution. Figure 2 shows the scatter diagram of the node degree and the frequency of five networks, including Karate, Dolphin and so on. The size of Data in Karate and Dolphins is smaller and the data distribution is discrete, most of the nodes have lower degrees, and only a few nodes have higher degrees, which is consistent with power-law distribution.

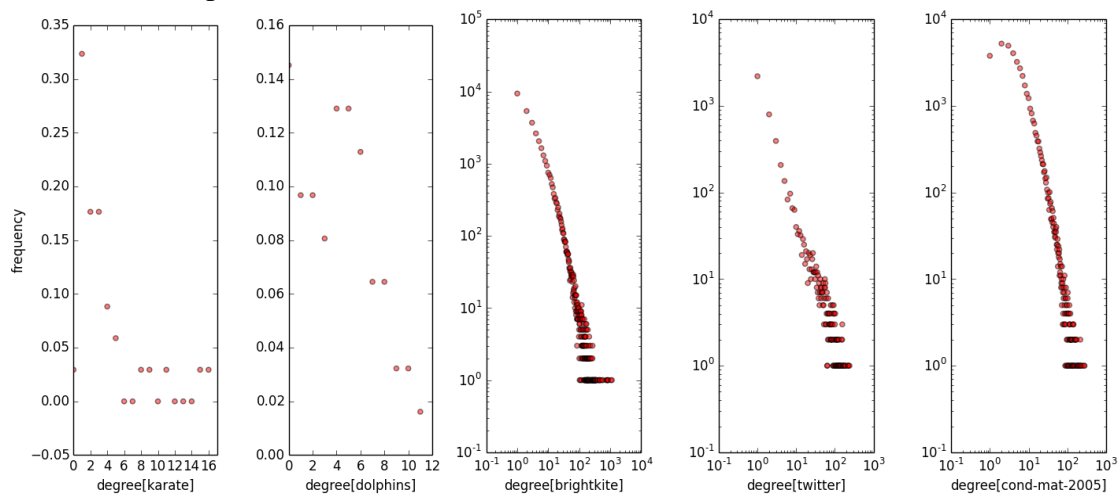

Fig.2.The scatter diagram *of node degree*

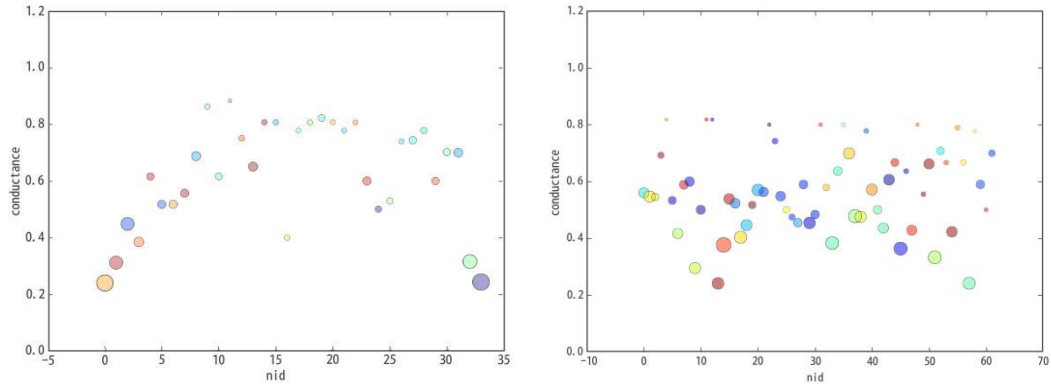

Fig.3.The karate and dolphins Network node conduction scatter diagram

Fig.3 shows the karate and dolphins Network node conduction, the horizontal axis denotes the nodes number, the vertical axis denotes the corresponding value of conductivity, and the area of scattered points relies on the degree. In the karate network, it is obvious that the maximum degree and minimum conduction prove the community structure, and node 0 and node3 could explain the situation. Meanwhile, it shows that the relationship between conductivity and degree is directly proportional in the strong community network, and both of them can be used to evaluate the importance of nodes. However, in the Dolphins network, there is no significant positive correlation between the node's degree and the node's conductivity. The node20 has a higher degree, when it has higher conductivity, which shows that the social circles with a large degree node may not necessarily have the obvious community characteristics. Therefore, the higher degree of a node may not necessarily lead to a higher quality of a community.

Fig.4 shows the logarithmic distribution of the conductivity in Cond-mat scientific collaboration network, the Twitter network, and the Brightkite network. It shows that the degrees of nodes between 10 and 100 have lower conductivity values. If the degrees of nodes are large, but the conductivity value of its neighbor set is small, then it demonstrates the neighbor set with large degree gets less difference between the closeness of internal neighbor and the external neighbor set, and that is to say, the community structure is not obvious. Thus, selecting the seed that depends on the node degree will lead to a poor quality of the community division.

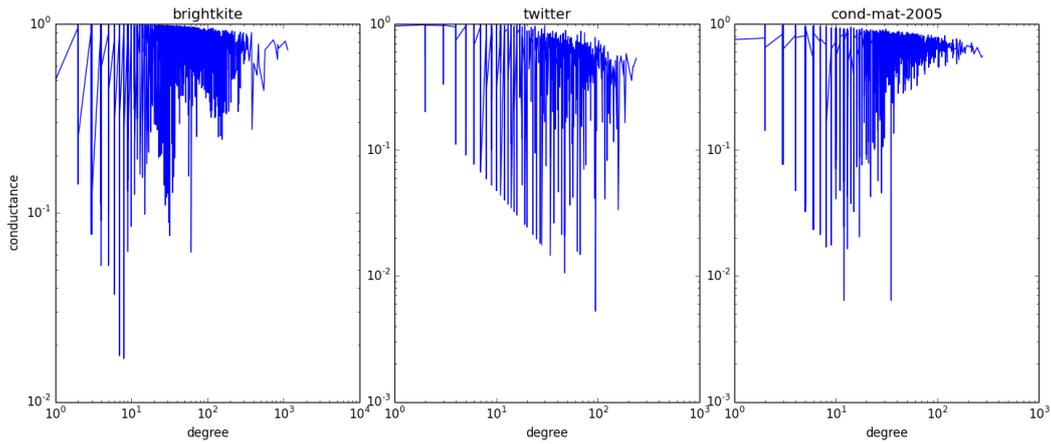

Fig.4.The conductivity of neighbor sets

In this paper, we choose FastUnfolding, CNM, LPA and Infomap algorithm for comparisons by the parameters of time complexity and the modularity. Table 2 shows the modularity about the NCB algorithm and the four algorithms on the Karate, Football and Dolphins data sets. FastUnfolding, CNM are the optimization algorithm based on the modularity, so the two algorithms achieve a high degree of modularity. And the result of the LPA algorithm is not unique, so we use the average degree of modularity which runs the LPA algorithm on three data sets for 5 times. The results of the LPA algorithm have high fluctuations. Infomap algorithm uses Huffman encoding by random walk for community structure detection, it can find a better community in the network if the community structure is obviously; and the NCB algorithm cannot produce the highest modularity, but the results perform closely to the FastUnfolding algorithm.

Table 2 The results of community division

| Algorithm | Karate | Football | Dolphins |
| --- | --- | --- | --- |
| CNM | 0.381 | 0.550 | 0.495 |
| LPA | 0.345[0.132,0.402] | 0.581[0.563,0.602] | 0.458[0.373,0.502] |
| Infomap | 0.402 | 0.600 | 0.528 |
| FastUnfolding | 0.419 | 0.605 | 0.519 |
| NCB | **0.378** | **0.585** | **0.510** |

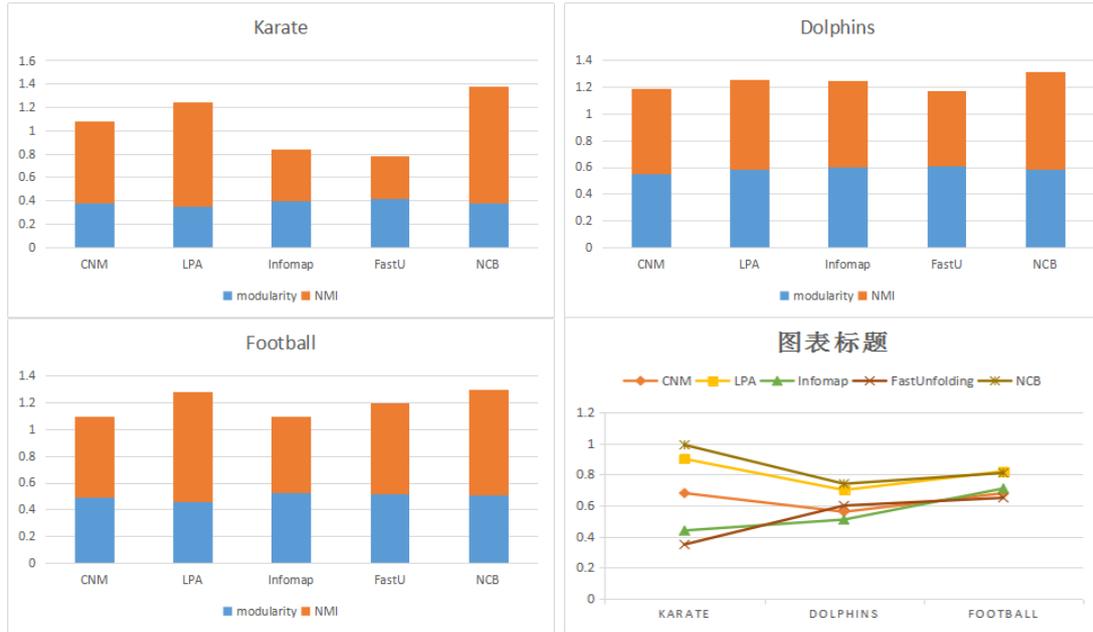

Fig.5 Comparison of NMI

In this paper, we use normalized mutual information to evaluate the algorithm, Since the result of LPA algorithm is not stable, we will set the maximum value of the 5 operations. As shown in Figure 5 (NMI), the correct rate of the LPA algorithm and the NCB algorithm are the best, the accuracy rate of NCB algorithm on Karate data set is 100%, the second is LPA, the accuracy of the other three algorithms are below 80%.

According to the NMI value, the FastUnfolding, CNM, and Infomap algorithms have achieved larger Q values but the accuracy are lower than the results with LPA and NCB. This shows a large difference between the cluster structure and the topology structure in the datasets with real background. In Figure 5, the overall score and performance of NCB algorithm are the best.

Table 3 Compare the number of communities found by different algorithms on the three datasets.

| Algorithm | Cond-mat | Twitter | Brightkite |
| --- | --- | --- | --- |
| CNM | 1910 | 168 | 1034 |
| LPA | 3590 | 648 | 1569 |
| Infomap | 3233 | 607 | 4829 |
| FastUnfolding | 1667 | 136 | 951 |
| NCB | 2267 | 366 | 1260 |

We use CNM, LPA, Infomap, FastUnfolding and NCB algorithms to do the experiments with three datasets (Cond-Mat-2005, Twitter, Brightkite) respectively, and try to calculate the number of communities, the degree of modularity of the community, and the time it takes (in seconds). Since the results of LPA are not unique, the results of LPA are the average value over the five times.

Table 3 lists the communities numbers found by each algorithm. LPA algorithm is a label-based diffusion algorithm, so in the larger network, the smaller the number of iterations can get the results. Infomap algorithm takes random walk to get the optimization function, so the community size are small and the number of communities is large. As shown in Table 3, the number of community found by Infomap and LPA algorithm is more than twice as much as the number of community found by other three algorithms on the three data sets. This indicates that LPA and Infomap algorithms tend to find smaller communities when the network size is large. CNM and FastUnfolding algorithms are modularity optimization algorithm, so the number of communities found by them is closer. And the NCB algorithm is a heuristic algorithm based on a local best community, so it can both ensures the discovery of larger communities and prevents the small communities to annexe by large communities. Therefore, the number of communities found by the NCB algorithm is between the results of the two algorithms described above, and the number of communities found by the NCB algorithm on Cond-mat-2005, Twitter, and Brightkite datasets are 2267, 366, and 1560, respectively.

On the Cond-mat-2005 dataset, there are not many differences in modularity for the community partitioning by each algorithm, but it can be seen that the modularity of CNM, FastUnfolding and NCB are slightly higher than that of LPA and Informap, as shown in Table 4.

Table 4 Comparison with the Modularity

| Algorithm | Cond-mat | Twitter | Brightkite |
| --- | --- | --- | --- |
| CNM | 0.679 | 0.869 | 0.603 |
| LPA | 0.662 | 0.794 | 0.455 |
| Infomap | 0.674 | 0.825 | 0.581 |
| FastUnfolding | 0.722 | 0.896 | 0.664 |
| NCB | **0.681** | **0.826** | **0.611** |

In Table 5, the theoretical time complexity of the five algorithms and their actual time spent on community discovery on three datasets are presented. Brightkite has 58,228 nodes and 214,078 edges, so the algorithm takes the longest time on this data set, and the shortest time on the Twitter. The Infomap gets the highest time complexity and the actual execution time is also the longest, the time is 869.767s; And the time complexity of NCB, LPA and FastUnfolding are all o(n), the actual lengths of execution time are different. The running time of FastUnfolding algorithm is the shortest, the efficiency is obviously better than the other four algorithms.

Table 5  Community partitioning results(time s)

| Algorithm | Cond-mat | Twitter | Brightkite |
|---|---|---|---|
| CNM | 250.7 | 68.15 | 358.88 |
| LPA | 72.40 | 49.74 | 151.63 |
| Infomap | 639.766 | 51.663 | 869.767 |
| FastUnfolding | 45.39 | 18.79 | 127.60 |
| NCB | 56.19 | 23.64 | 161.30 |

Based on the three kinds of statistical data, the Fast Unfolding algorithm is the best in terms of time and modularity, but it is less effective for small-grained communities and has the least number of communities found. The NCB algorithm has less complexity in time complexity than FastUnfolding algorithm, and the modularity of community partitioning is closer to FastUnfolding, but NCB algorithm finds more communities than FastUnfolding algorithm, so NCB algorithm can be also applied to small-grained communities.

## 4 Conclusion

In this paper, we have proposed a kind of community discovery algorithm based on local core members (NCB algorithm) to solve the problem of community detection in social networks. The NCB algorithm calculates the local core nodes in the network via computing the conductivity values. Then, through using the neighbor sets, the core nodes are processed to form the initial community. Finally, the nodes of the unmarked community are updated iteratively. The conductivity value of the community determines the community ownership. Experiments on Karate and Dolphins networks with real communities have shown that the community structure obtained by NCB algorithm is the closest to the true community structure, and the accuracy is the highest. In addition, although the NCB algorithm cannot produce the maximum Q value on Twitter and Cond-mat networks, the NCB algorithm can find small-sized and better-quality communities, and can complete the computation in a linear time complexity. Compared with other algorithms in terms of modularity, community number, running time and accuracy, NCB algorithm has the best overall performance. But we do not consider the influence of node attributes in the diffusion model, and will try to find such community structure in the future works.